# Lunar Exploration: Opening a Window into the History and Evolution of the Inner Solar System

Ian A. Crawford[1,2] and Katherine H. Joy[3]

[1] Department of Earth and Planetary Sciences, Birkbeck College, University of London, Malet Street, London, WC1E 7HX, UK.

[2] Centre for Planetary Sciences at UCL/Birkbeck, Gower Street, London, WC1E 6BT, UK.

[3] School of Earth, Atmospheric and Environmental Sciences, University of Manchester, Williamson Building, Oxford Road, Manchester, M13 9PL, UK.

## Abstract

The lunar geological record contains a rich archive of the history of the inner Solar System, including information relevant to understanding the origin and evolution of the Earth-Moon system, the geological evolution of rocky planets, and our local cosmic environment. This paper provides a brief review of lunar exploration to-date, and describes how future exploration initiatives will further advance our understanding of the origin and evolution of the Moon, the Earth-Moon system, and of the Solar System more generally. It is concluded that further advances will require the placing of new scientific instruments on, and the return of additional samples from, the lunar surface. Some of these scientific objectives can be achieved robotically, for example by *in situ* geochemical and geophysical measurements and through carefully targeted sample return missions. However, in the longer term, we argue that lunar science would greatly benefit from renewed human operations on the surface of the Moon, such as would be facilitated by implementing the recently proposed Global Exploration Roadmap.

## 1. Introduction

From a scientific perspective lunar exploration has advanced, and in the future has the potential to continue to advance, human knowledge in three broad areas. Firstly, the Moon preserves a record of the early geological evolution of a rocky planet (including planetary differentiation and magma ocean processes), which more evolved planetary bodies have largely lost, as well as geochemical and geophysical constraints on the origin and evolution of the Earth-Moon system [1-4]. Secondly, the lunar surface, and especially the lunar regolith, contains records of inner Solar System processes (e.g., meteorite flux, interplanetary dust density, solar wind flux and composition, and galactic cosmic ray flux) throughout most of Solar System history, much of which is relevant to understanding the history and evolution of our own planet and its biosphere [1,4-7]. Thirdly, the lunar surface is a potential platform for a range of scientific investigations, notably observational astronomy [8,9] (especially low frequency radio astronomy from the far-side [10]), but possibly in the future also extending to investigations in fundamental physics [11], astrobiology [12], and human physiology and medicine [13].

In this paper we first give a brief historical summary of lunar exploration to-date, and then discuss how future lunar exploration can contribute to the development of the first two of the three broad scientific fields outlined above. The third scientific area discussed above, namely the potential value of the Moon as a platform for astronomical and other scientific investigations, although likely to be an important part of future lunar exploration activities, lies outside the scope of the meeting reported here (readers interested in those aspects of lunar exploration are instead referred to the reviews in [8-14] and references cited therein).

## 2. A brief history of lunar exploration

The modern scientific investigation of the Moon as a planetary body began with Galileo's first telescopic observations in 1609 [15], and telescopic observations of the near-side have continued ever since [16]. However, the bulk of our knowledge of lunar geological evolution, and its implications for Solar System history as a whole, has been obtained through direct investigation by space probes only during the last half-century or so [17,18]. Table 1 provides a summary of the most important spacecraft to have visited the Moon as of April 2014.

The first spacecraft to reach the Moon was the Soviet Union's Luna 2 spacecraft, which impacted the lunar surface on the 13th September 1959. Of greater significance for lunar geology was the flight of Luna 3, in October that same year, which completed the first flyby of the Moon and obtained the first ever images of the lunar far-side, revealing that the far-side is largely devoid of the dark expanses of basaltic lava that dominate the near-side. After a short break of six years, Luna 9 successfully soft-landed and obtained the first surface images in February 1966, and Luna 10 became the first spacecraft to enter orbit about the Moon in April of the same year.

During this period the US lunar exploration programme started ramping up in response to President Kennedy's initiation of the Apollo programme in May 1961. The first US lunar probes were the Ranger series of 'hard landers,' designed to take ever increasing resolution images of the surface before crashing into it, which paved the way for the Surveyor series of robotic soft landers between 1966 and 1968 (Fig. 1). In parallel, between 1966 and 1967 the US flew a highly successful series of Lunar Orbiter spacecraft that were designed to obtain high resolution images of the lunar surface. With surface resolutions of several tens of metres (occasionally as high as 2 m), these images long remained unsurpassed as a resource for lunar geology (although they are now rapidly being superseded by images obtained by the Lunar

Reconnaissance Orbiter Narrow Angle Camera discussed below). In large part, the Lunar Orbiter missions were designed to identify potential landing sites for the manned Apollo missions then under development, just as the Surveyors were designed to provide knowledge of the surface environment with the manned landings in mind.

The Apollo programme is of pivotal importance in the history of lunar exploration, and it has left an enduring scientific legacy [e.g. 17-20]. Between July 1969 and December 1972 a total of twelve astronauts explored the lunar surface in the vicinity of six Apollo landing sites (Figs. 1,2). The total cumulative time spent on the lunar surface was 25 person-days, with just 6.8 man-days spent performing exploration activities outside the lunar modules. Over the six missions, the astronauts traversed a total distance of 95.5 km from their landing sites (heavily weighted to the last three missions that were equipped with the Lunar Roving Vehicle), collected and returned to Earth 382 kg of rock and soil samples (from over 2000 discrete sampling localities), drilled three sample cores to depths of 2-3 m, obtained over 6000 surface images, and deployed over 2100 kg of scientific equipment (including seismometers heat-flow probes, magnetometers, gravimeters, and laser-ranging reflectors [17-20]). These surface experiments were supplemented by remote-sensing observations conducted from the orbiting Command and Service Modules.

Two important Soviet robotic programmes overlapped with Apollo, and continued to keep lunar surface exploration alive for a few years after human exploration ceased. These were the two 'Lunokhod' rovers (Luna 17 and 21) that landed on the Moon in 1970 and 1973, and the three robotic sample return missions (Luna 16, 20 and 24) of 1970, 1972, and 1976, respectively. The Lunokhods were the first tele-operated robotic rovers to operate on another planetary body. Lunokhod 1 operated for 322 days and traversed a total distance of 10.5 km in the Sinus Iridum; the corresponding numbers for Lunokhod 2 were 115 days and 37 km in Le Monnier crater on the edge of Mare Serenitatis [21]. During their traverses the Lunokhods made measurements of the regolith's mechanical properties (using a penetrometer) and composition (determined using a X-ray fluorescence spectrometer), as well as the surface radiation environment; they also carried reflectors which, similar to those deployed by the Apollo 11, 14 and 15 missions, have been used to measure the Earth-Moon distance and the Moon's physical librations. The Luna 16, 20 and 24 missions collected, and returned to Earth, a total of ~320 grams of lunar soils from three sites close to the eastern limb of the near-side (Fig. 2). Although the quantity of material collected was small compared to that returned by Apollo, their geographical separation from the Apollo landing sites makes the Luna samples important for our understanding of lunar geological diversity.

Following the Luna 24 mission in 1976 there was almost a twenty-year gap in lunar exploration, only broken in the 1990's when the Hiten, Clementine and Lunar Prospector spacecraft flew to the Moon and heralded a renewed era of lunar exploration (Table 1). Although pioneering from a space technology standpoint [22], the Japanese Hiten probe and its associated dust detection instrument did not reveal significant new information about the Moon. On the other hand, the Clementine [23] and Lunar Prospector [24] orbital missions proved crucial by providing global mineralogical and geochemical maps of the lunar surface. Data obtained by these two missions clearly showed that the lunar surface is geologically much more diverse than had been suspected based on the Apollo and Luna samples, and stimulated renewed scientific interest in the geological evolution of the Moon and its implications for planetary science more widely [25].

Partly as a result of this renewed scientific interest in the Moon, and partly as a result of emerging space powers wishing to show-case newly acquired technical expertise, the last decade has seen a renaissance in lunar exploration conducted from orbit. In the last ten years the following countries have all sent remote-sensing spacecraft to lunar orbit (Table 1): European Space Agency: SMART-1 (2003 [26]); Japan: Kaguya (2007 [27]); China: Chang'e-1 (2007 [28]), Chang'e-2 (2010 [29]); India: Chandrayaan-1 (2008 [30]); and the United States: Lunar Reconnaissance Orbiter (LRO; 2009 [31]), GRAIL (2012 [32]), and LADEE (2013 [33]). This plethora of orbital missions has added significantly to our knowledge of the lunar surface and, in the case of Kaguya and GRAIL, to the lunar interior. However, it is notable that none of these spacecraft were designed to land on the Moon's surface in a controlled manner (although the US Lunar CRater Observation and Sensing Satellite (LCROSS [34]; co-launched with LRO) and the Chandrayaan-1 Moon Impact Probe (MIP [35]) did deliberately impact the lunar surface in an effort to detect polar volatiles).

Most recently, in December 2013, China successfully landed the Chang'e-3 vehicle, equipped with a small rover, Yutu, in northern Mare Imbrium. This achievement has broken a 37-year hiatus in lunar surface exploration, being the first controlled soft landing on the Moon since the Soviet robotic sample return mission Luna 24 in August 1976. Among other experiments, Yutu carried a ground penetrating radar instrument to study sub-surface regolith structure [36], which is the first time such an instrument has been deployed on the lunar surface.

## 3. Current plans for near-term future lunar exploration

In the 2014-2022 timeframe there are tentative plans for a number of other robotic landings on the lunar surface, although most remain unconfirmed and precise timings are uncertain.

Building on the success of Chang'e-3, China is likely to land Chang'e-4 and Chang'e-5 in or around 2015 and 2017, respectively. Depending on the success of the earlier missions, Chang'e-5 is intended to be a sample return mission (the first since Luna 24 in 1976), although its landing site has not yet been determined [37,38]. In the same timeframe Japan is likely to deploy its dual orbiter/rover Selene-2 mission [39], and India has stated an intention to return to the Moon with its proposed Chandrayaan-2 mission [40].

Russia has plans for an increasingly sophisticated set of orbiters and landers (to be named Luna 25-28) in the period 2016-2021, of which Luna 28 is planned to be a sample return mission from a near-polar locality [41]. In the 2018-20 time-frame it appears likely, although not yet confirmed, that the US Resource Prospector Mission with its rover-borne RESOLVE ('Regolith and Environment Science and Oxygen and Lunar Volatile Extraction') payload [42] will be deployed to investigate high-latitude lunar volatile deposits. Moreover, towards the end of this decade NASA aims to deploy its manned 'Orion' Crew Exploration Vehicle to the second Earth-Moon Lagrange point, which, although it will not itself provide access to the surface, may nevertheless facilitate far-side surface exploration by acting as a communications relay and as a node for the tele-operation of surface instruments [43].

Over the last ten years a large number of additional lunar mission studies have been conducted, some of which are discussed in Section 4 below, but to-date none have received funding or space agency support. It is also worth noting that, in addition to government-led activities, the coming decade may also witness privately funded lunar landings, conducted in pursuit of the Google Lunar X-Prize [44] or other private initiatives, although the scientific opportunities presented by these relatively small missions remain to be determined.

## 4. Scientific objectives for future lunar exploration

A careful, top-level, prioritisation of lunar science objectives was given by a US National Research Council (NRC) study in 2007 [1], and this is summarised in Table 2. Addressing most of these questions satisfactorily will not be achieved by further orbital remote-sensing missions but will require the return of additional samples from, and the placing of a new generation of scientific instruments on, the surface of the Moon [1,5,14,45,46]. The NRC scientific prioritisation continues to represent a consensus among the lunar science community, and forms the basis for planning future lunar exploration strategies [e.g. 14,45,46], including detailed landing site assessment studies [47,48]. Rather than attempt to reiterate the results of these previous studies here, we instead focus on those aspects of lunar exploration which will advance our knowledge of the origin and early evolution of the Moon and of inner Solar System history more generally. These exploration objectives all map onto the top-level NRC science questions listed in Table 2 [1], but considering them under these two broad themes better emphasises the synergies between them and how lunar exploration can contribute to the central topic discussed in this volume.

### (a) Understanding the origin and early evolution of the Earth-Moon system

Other papers in this volume amply demonstrate the value of past lunar exploration, and especially the Apollo and Luna samples, in constraining theories of the Moon's origin and evolution. It is clear that without access to these lunar materials our understanding of the origin of the Earth-Moon system would be even less complete than it currently is. In terms of future exploration objectives it is possible to identify several high-priority areas for *in situ* geophysical and geochemical measurements and/or sample return locations.

(i) Surface geophysical measurements

The interior of the Moon is expected to retain a record of early planetary differentiation processes that more evolved planetary bodies have since lost [1,3,14]. In the present context, improved knowledge of the lunar interior will inform models of the Moon's early thermal state, including those related to magma ocean formation and evolution (Fig. 3), which will be helpful in constraining theories of the Moon's origin and earliest evolution [49]. Obtaining such information will require making further geophysical measurements. While some relevant measurements can be made from orbit, such as the measurement of the Moon's gravity [32,50] and magnetic [51] fields, most will require geophysical instruments to be placed on, or just below, the lunar surface. Key instruments in this respect are seismometers, to probe the structure of the deep interior [52,53], heat-flow probes to measure the heat loss from the lunar interior and its spatial variations [54], and magnetometers to determine interior electrical and magnetic properties from induced magnetic fields [55]. Such instruments could be placed and operated on the lunar surface by robotic landers, such as envisaged for the proposed Farside Explorer [56], Lunette [57], and LunarNet [58] mission concepts.

(ii) Samples of the lunar mantle

Models of the Moon's formation rely heavily on constraints provided by the bulk chemical and isotopic compositions of terrestrial and lunar rocks [59-63]. However, while we have relatively good knowledge of the composition of the bulk silicate Earth (obtained from direct measurements of mantle xenoliths and from modelling the composition of countless lava flows derived from the mantle by partial melting [64,65], our knowledge of the composition of the bulk silicate Moon is far less secure [66]. To-date, no samples of the lunar mantle have been identified, and inferences about the mantle composition based on reconstructions from partial melt compositions are constrained by the limited range of lunar samples in the Apollo and Luna sample collections.

Geochemical and isotopic inferences regarding the origin and evolution of the Moon would be more robust if we had samples of the lunar mantle to study. There are several ways in which this might be achieved, but all involve returning to the Moon to obtain additional samples. One possibility would be to target a sample return mission to a locality where orbital remote-sensing data indicates that mantle materials may be exposed at the surface, mainly around the periphery of large impact basins [67]. Of particular interest are areas where the gravity data indicate a very low, or even non-existent, crustal thickness [68]. One such locality is the Crisium basin, where the two ~20-km diameter craters Peirce and Picard may have penetrated through a relatively thin overlying basaltic fill to excavate underlying mantle material [3,69].

Another possibility would be to search for mantle xenoliths within lunar basalts. Although, to our knowledge, no xenoliths have been identified in Apollo basalt samples, a more extensive sampling of lunar lava flows might discover them. It would be surprising if no lunar lavas ever entrained mantle xenoliths, which are often common in terrestrial basaltic lava flows. Such samples would not only allow direct bulk and isotopic measurements to be made of the mantle materials, but, even if they turn out to be very rare, a comparison between their compositions and the encapsulating basalt would permit tests of partial melting scenarios used to backtrack from basalt to mantle compositions from samples lacking xenoliths. Moreover, even in the absence of finding mantle xenoliths, a wider sampling of lunar basalts is in any case desirable for studies of lunar mantle composition and evolution just because of the geographically limited range of the existing samples.

(iii) Samples of the lunar highlands

Early theories born out of the analyses of Apollo samples favoured the idea that rapid accretion of the Moon gave rise to a lunar magma ocean (LMO) that encompassed the whole or a substantial part of the lunar interior [49, 70]. However, the global extent and timing of crustal formation episode(s) is now under debate (Fig. 3), with important ramifications for understanding the duration and magnitude of the LMO, and thus the timing of the Moon forming event itself [71]. This relatively recent controversy has arisen from two main lines of evidence.

Firstly, remote sensing datasets have revealed that the Apollo sample collection is largely derived from a geochemically anomalous region (the 'Procellarum KREEP Terrain' (PKT) [72]), and may not be truly representative of the global lunar highlands crust. This is supported by the availably of new samples collected on Earth as lunar meteorites, many of which may have originated from crustal regions remote to the nearside geochemical anomaly [73]. Significant chemical differences are found between feldspathic lunar meteorites and near-side highland ferroan anorthosites (FAN), which implies a degree of compositional heterogeneity among highland lithologies that cannot easily be accounted for in the standard LMO paradigm [74-78].

Secondly, advances in the field of geochronology have revealed an apparent overlap in the timing of feldspathic crust formation and magmatic episodic intrusions into this crust at ~4.3-4.2 Ga [71,76,79], challenging the view that the Moon had differentiated completely by ~4.4 Ga. In particular, neodynium isotopic compositions of samples from the lunar crust (both Apollo and lunar meteorite samples) require several distinct geochemical source regions [71, 80]. This suggests that the crust may not have formed in a simplistic single magma ocean floataation event, and that more complex and multi-scale geological processes (e.g. regional magma ocean/seas, serial magmatism, and/or large-scale differentiated impact melt sheets; Fig. 3) may be involved in its formation.

To effectively address questions about the age and diversity of the lunar highlands crust, future sample return missions should focus on collecting material from regions of the Moon remote from where the Apollo samples were obtained (i.e., from the far-side of the Moon, the polar regions, and the southern nearside highlands). Moreover, the interpretation of existing samples is compromised by a lack of knowledge of their bedrock sources (for example, the Apollo samples were retrieved from regoliths developed on top of basin ejecta sourced from varying crustal depths, and lunar meteorites have poorly spatially constrained launch localities), so future sampling efforts should focus on retrieving material of known geological and stratigraphic provenance. For example, the uplifted central peaks and peak rings of lunar basins and large craters provide access to material excavated from different depths [81]. Thus, direct sampling of central peaks in a range of different size craters would provide samples from the upper to lower lunar crust. Orbital remote-sensing has identified several such regions as having outcrops of essentially pure anorthosite [82] that might best represent pristine samples of LMO derived floataation crust(s). Accessing such regions would be of great interest, but would require precision landing coupled with rover-facilitated mobility in addition to a sample return capability.

It is also important to realise that, despite the increasing sensitivity of analytical techniques, the interpretation of isotope data for highland samples is often limited by the small sample masses available, owing to the relative rarity of suitable minerals to provide robust isotopic age dates (e.g. [71, 79]). For example, although Borg et al. [79] conducted their dating of a ferroan anorthosite sample having a mass of only 1.9 g, in order to identify a sufficiently mafic-rich (~25% pyroxene) anorthositic clast suitable for analysis they had first to conduct a thorough search of the extensive Apollo sample collection. It follows that high-precision isotopic studies of the lunar highlands aimed at addressing the outstanding questions relating to lunar crustal evolution will require significant masses (at least tens of grams, and perhaps even kilograms if multiple analyses are required) of anorthositic highland rocks in order to obtain appreciable quantities of datable mineral phases. Although future robotic sample return missions to suitably chosen locations would help address these issues, retrieving the large sample masses required, and the implied surface mobility requirements, would be greatly facilitated by future human surface exploration initiatives (see discussion in [14]).

(iv) Samples not contaminated by cosmic rays and the solar wind

The lunar surface is the interface between the Moon and the surrounding space environment [6,7]. As such it is continually bombarded by solar wind particles and galactic cosmic rays. Spallation reactions occur when an incident cosmic ray (proton, neutron or alpha particle) hits a target element with energies that are high enough to produce a variety of radiogenic and stable 'cosmogenic' nuclides [83,84]. The resultant nuclides are useful as they can be measured to calculate the time duration to space exposure (based on nuclide production rates for the specific sample chemistry). However, they also complicate the determination of primordial isotopic signatures, especially as lighter isotopes are affected by this spallation process more than the heavier isotopes, creating apparent fractionated isotope ratios. Knowledge of cosmogenic spallation processes, thus, requires a correction to be applied to measured isotope ratios and abundances, which is dependent on prior knowledge of the sample's exposure record and chemistry.

This effect is particularly relevant for key isotope systems used to investigate chemical processes in the giant impact event and in lunar differentiation. For example, in principle hydrogen isotope abundances can provide key insights to the nature of volatile fractionation during impact events, but as deuterium (D) can be produced cosmogenically the measured D/H ratios must be corrected for space exposure effects before they can be properly interpreted [85]. Similarly, hafnium (Hf) and tungsten (W) isotopes provide constraints on the timing of core formation, providing important information about the age of the Moon and the giant impact event, but the isotope $^{182}$W can be generated by spallation processes and its measured abundance must corrected to account for this effect [60]. Thus, to some extent the interpretation of these important isotope systems is limited by our knowledge of the space exposure records and cosmogenic nuclide production in the samples.

It is important to note is that spallation reactions are limited to the upper few metres (~5 m) of the lunar regolith, as incident particles cannot penetrate to greater depths [84]. Thus samples obtained from greater depth will not have experienced space exposure and will preserve their original isotopic records. Future lunar sampling efforts should therefore include the collection of material that has never been exposed to cosmic rays or the solar wind, or at least has only been exposed for geologically insignificant durations. Examples might include crustal material that has been excavated by impact craters (provided that samples can be obtained from below the surficial regolith covering), and buried lava flows and pyroclastic deposits that have been rapidly covered by younger lava flows more than several metres thick. Examples of the latter appear to be common in the lunar maria (Fig. 4), but accessing them will require sufficient mobility to access suitable outcrops and/or the development of a drilling capability to retrieve material from depths of several metres.

## (b) Accessing lunar records of Solar System history

In addition to providing information relevant to understanding the origin and earliest evolution of the Earth-Moon system, a vigorous lunar exploration programme would greatly add to our knowledge of the inner Solar System, and especially the near-Earth, cosmic environment throughout most of Solar System history. This is because the lunar regolith contains a record of the inner Solar System fluxes, and to a degree also the changing compositions, of asteroids, comets, interplanetary dust particles, solar wind particles, and galactic cosmic rays over at least the last 4 Ga [e.g. 1,5-7,14,86-89]. More speculatively, the lunar regolith may also contain samples of Earth's earliest crust in the form of terrestrial meteorites [90] and atmosphere [91] not otherwise available. We here elaborate on those aspects of the lunar geological record most relevant to understanding the history of the near-Earth environment, much of it relevant to understanding the past habitability of our own planet, and how future lunar exploration may best address them.

(i) The Impact History of the Inner Solar System

Most lunar surfaces have never been directly dated, and their inferred ages are based on the observed density of impact craters calibrated using the ages of Apollo and Luna samples [87]. However, this calibration, which is used to convert crater densities to absolute model ages, is neither as complete nor as reliable as is often supposed. As a result of the limited age range of surfaces sampled by the Apollo and Luna missions, there are no calibration points older than about 3.85 Ga, and crater ages younger than about 3 Ga are also uncertain [92,93]. Improved knowledge of the lunar cratering rate would be of great value for planetary science for at least three reasons: (i) it would yield improved estimates for the ages of lunar terrains for which samples are not yet available; (ii) it would result in a more complete knowledge of the impact history of the inner Solar System, including that of Earth; and (iii) because the lunar impact rate is used, with various assumptions, to date surfaces on planets, moons and asteroids for which samples have not been obtained (and in some cases may never be obtained) uncertainties in the lunar impact chronology result in uncertainties in the ages of planetary surfaces throughout the Solar System.

An important unresolved question is whether the inner Solar System cratering rate has declined monotonically since the formation of the Solar System, or whether there was a bombardment 'cataclysm' between about 3.8 and 4.1 Ga ago characterised by an enhanced rate of impacts (Fig. 5) [87, 94-96]. Indeed, recent studies of the ages of impact melt samples obtained by the Apollo and Luna missions suggest a very complicated impact history for the Earth-Moon system, with a number of discrete spikes in the impact flux [97]. Clarifying this issue is especially important in an astrobiology context as it defines the impact regime under which life on Earth became established and the rate at which volatiles and organic materials were delivered to the early Earth [94, 98-100]. Additionally, as the inner Solar System bombardment history is thought to have been governed, at least in part, by changing tidal resonances in the asteroid belt [96, 101-103] improved constraints on the impact rate will lead to a better understanding of the orbital evolution of the early Solar System.

Obtaining an improved lunar cratering chronology requires the radiometric dating of surfaces having a wide range of crater densities, supplemented where possible by dating of impact melt deposits from individual craters and basins [87,97]. In practice this will require either *in situ* dating or sample return missions to specially chosen localities. Farley et al. [104] have recently demonstrated the *in situ* radiometric dating of rocks on Mars using instruments on the *Curiosity* rover and, although such robotic techniques are never likely to be as accurate as measurements made in terrestrial laboratories, they could nevertheless be extremely valuable if applied to areas from which samples have not yet been obtained. Developing an *in situ* dating technique suitable for robotic landers and rovers would therefore be a very useful addition to the tools of lunar geochronology. That said, for the foreseeable future, it seems certain that the most accurate dating techniques will continue to require samples to be returned to Earth for analysis.

Key lunar sampling sites for characterising the inner Solar System impact history include the far-side South Pole-Aitken basin (the dating of which will help determine the duration of the basin-forming epoch [e.g. 94,96,105]), the near-side Nectaris basin (a key lunar stratigraphic marker, the age of which would help determine the existence and duration of a hypothesised late spike in the rate of basin-forming impacts [87; A. Morbidelli, personal communication), and, at the other end of the age spectrum, young basaltic lava flows in Oceanus Procellarum on the near-side (where the dating of individual lava flows with ages in the range 1.1 to 3.5 Ga would provide data points for the as yet uncalibrated 'recent' portion of the inner Solar System cratering rate [4,87,106,107]; Fig. 6).

(ii) Treasures in the regolith

The lunar regolith is known to contain much that is of interest for studies of Solar System history. For example, studies of Apollo samples have revealed that solar wind particles are efficiently implanted in the lunar regolith [6,7], which therefore contains a record of the composition and evolution of the Sun [108-110]. Recently, samples of the Earth's early atmosphere may have been retrieved from lunar regolith samples [91], and it has been suggested that samples of Earth's early crust may also be preserved there in the form of terrestrial meteorites [90, 111-113]. Meteorites derived from elsewhere in the Solar System have already been found on the Moon, preserving a record of the dynamical evolution of small bodies throughout Solar System history [88]. In addition, the lunar regolith may contain a record of galactic events, by preserving the signatures of ancient galactic cosmic ray fluxes, and the possible accumulation of interstellar dust particles during passages of the Sun through dense interstellar clouds [6,114,115]. Collectively, these lunar geological records would provide a window into the early evolution of the Sun and Earth, and of the changing galactic environment of the Solar System, that is unlikely to be obtained in any other way.

From the point of view of accessing ancient Solar System history it will be especially desirable to find layers of ancient regoliths ('palaeoregoliths') that were formed and buried long ago, thereby ensuring that the records they contain come from tightly constrained time horizons (Fig. 7) [5,107, 115-117]. Locating and sampling such deposits will therefore be an important scientific objective of future lunar exploration activities [106], but they will not be easy to access. Although robotic sampling missions might in principle be able to access palaeoregoliths at a limited number of favourable sites (for example where buried layers outcrop in the walls of craters or rilles), fully sampling this potentially rich archive of Solar System history will probably require the mobility, drilling, and sample return capabilities of future human exploration [5, 14, 107, 118].

(iii) Polar volatiles

The lunar poles potentially record the flux of volatiles present in the inner Solar System throughout much of Solar System history [1]. The Lunar Prospector neutron spectrometer found evidence of enhanced concentrations of hydrogen in the regolith near the lunar poles [119], an observation that was widely interpreted as indicating the presence of water ice in the floors of permanently shadowed polar craters. This interpretation was supported by the LCROSS impact experiment, which found a water ice concentration of $5.6 \pm 2.9$ % by weight in the target regolith at the Cabeus crater [34]. This water is probably ultimately derived from the impacts of comets and/or hydrated meteorites on to the lunar surface [120]. In addition to possible ice in polar craters, infra-red remote-sensing observations have found evidence for hydrated minerals, and/or adsorbed water or hydroxyl molecules, over large areas of the high latitude (but not permanently shadowed) lunar surface which may be due to oxidation of solar wind hydrogen within the regolith [121,122].

Obtaining improved knowledge of the presence, composition, and abundance of water (and other volatile species) at the lunar poles is important for several reasons. Firstly, even though the original cometary and/or meteoritic volatiles will have been considerably reworked, it remains probable that some information concerning the composition of the original sources will remain [122]; this may yield important knowledge on the role of comets and meteorites in delivering volatiles and pre-biotic organic materials to the terrestrial planets [123-125]. Secondly, lunar polar ice deposits will have been continuously subject to irradiation by galactic cosmic rays and, as such, may be expected to undergo prebiotic organic synthesis reactions of astrobiological interest [126]. Thirdly, the presence of water ice at the lunar poles, and even hydrated materials at high-latitude but non-shadowed localities, could potentially provide a very valuable resource (e.g., rocket fuel, habitation resources) in the context of future lunar exploration activities [127].

Confirming the existence, abundance, and physical and chemical state of polar and high-latitude lunar volatiles will require in situ measurements by suitably instrumented spacecraft. Near-term opportunities, both proposed for the 2018-2020 timeframe, include the US Resource Prospector Mission [42] and Russia's proposed Luna 28 lunar polar sample return mission [41]. Penetrator-based concepts, such as proposed for MoonLITE [128] and LunarNET [58], also appear to be well adapted for characterising volatiles in permanently shadowed polar regions. Non-permanently shadowed high-latitude localities are of course amenable to solar-powered robotic exploration, and proposals such as 'Lunar Beagle' [129] would provide valuable initial measurements of volatiles in these environments. In the longer term, a full characterisation of polar volatiles, as for other aspects of lunar geology, would benefit from the increased mobility and flexibility that would be provided by human exploration (which would of course be facilitated at the poles if exploitable quantities of volatiles prove to be present [127,130]).

## 5. Conclusions

The lunar geological record contains a rich archive of the history of the inner Solar System, including information relevant to understanding the origin and evolution of the Earth-Moon system, the geological evolution of rocky planets, and our local cosmic environment. Gaining access to this archive will require a renewed commitment to lunar exploration, with the placing of a new generation of scientific instruments on, and the return of additional samples from, the surface of the Moon. Although robotic missions currently planned for the 2015-2020 time-frame will partially address some of these questions (Section 3), a much more ambitious programme of lunar exploration will be required in order to access the full potential of the lunar geological record. Robotic geophysics and sample return missions targeted at specific landing sites to address key outstanding questions would confer major scientific benefits. However, for many of the lunar exploration objectives that we have identified (Section 4), the requirements for mobility, deployment of complex instrumentation, sub-surface drilling, and sample return capacity are likely to outstrip the capabilities of robotic or tele-robotic exploration. It follows that many of these scientific objectives would be greatly facilitated by renewed human operations on the surface of the Moon, in much the same way, and for essentially the same reasons, as Antarctic science benefits from the infrastructure provided by scientific outposts on that continent [e.g. 5,14,45-48,118,131-136].

Recent developments in international space policy augur well for the initiation of such an expanded lunar exploration programme [136]. In particular, in 2007 fourteen of the world's space agencies produced the Global Exploration Strategy [137], which lays the foundations for an ambitious, global space exploration programme. One of the first concrete results of this activity has been the development of a Global Exploration Roadmap [138], which outlines possible international contributions to the human and robotic exploration of the inner Solar System over the next twenty-five years. This would provide many opportunities for pursuing the lunar science objectives outlined in this paper and, along with many other benefits to planetary science and astronomy, an increased understanding of the origin and evolution of the Earth-Moon system would be major scientific benefit of its implementation.

## Acknowledgements

We thank David Stevenson and Alex Halliday, organisers of the Royal Society meeting on the Origin of the Moon, for the invitation to present this paper. We further thank Alex Halliday for drawing our attention to the importance of obtaining lunar samples uncontaminated by cosmogenic isotopes, and Alessandro Morbidelli for reminding us of the importance of dating the Nectaris impact. We thank our many colleagues (especially Mahesh Anand, Neil Bowles, Ben Bussey, James Carpenter, Heino Falcke, Sarah Fagents, Ralf Jaumann, David Kring, Sara Russell and Mark Wieczorek, but also others too numerous to mention) for many helpful discussions and collaborations relating to lunar exploration, and

Wenzhe Fa for advice on the Chinese lunar exploration programme. Finally, we thank our three referees for helpful comments which have improved this paper. KHJ acknowledges Leverhulme Trust grant 2011-569.

**Table1**. Highlights of Lunar Exploration by Spacecraft. For programmes consisting of several spacecraft the numbers in parentheses donate the fraction of successful missions [18,21].

| Spacecraft/Programme Name | Nationality | Launch Year | Mission Description |
|---|---|---|---|
| Luna 2 | USSR | 1959 | First lunar impact |
| Luna 3 | USSR | 1959 | Flyby: first farside images |
| Ranger probes (3/7) | USA | 1962-65 | Impact probes: near surface imagery |
| Luna 9 | USSR | 1966 | Soft lander: first surface images |
| Luna 10 | USSR | 1966 | First lunar orbiter |
| Surveyor Landers (5/7) | USA | 1966-68 | Soft landers: surface properties |
| Lunar Orbiters (5/5) | USA | 1966-67 | Orbiters: orbital photography |
| Apollo 8, 10 | USA | 1968-69 | Manned lunar orbiters: orbital photography |
| Apollo 11, 12, 14-17 | USA | 1969-72 | Manned landings: surface and interior properties; sample return; orbital remote sensing |
| Lunokhod 1, 2 (Luna 17, 21) | USSR | 1970, 73 | Robotic rovers: surface properties |
| Luna 16, 20, 24 | USSR | 1970, 72, 76 | Robotic sample return |
| Hiten (MUSES-A) | Japan | 1990 | Lunar orbiter: dust detection |
| Clementine | USA | 1994 | Orbital remote sensing |
| Lunar Prospector | USA | 1998 | Orbital remote sensing |
| SMART-1 | Europe | 2003 | Orbital remote sensing |
| Kaguya | Japan | 2007 | Orbital remote sensing |
| Chang'e-1 | China | 2007 | Orbital remote sensing |
| Chandrayaan-1 | India | 2008 | Orbital remote sensing |
| Lunar Reconnaissance Orbiter | USA | 2009 | Orbital remote sensing |
| Lunar Crater Observation and Sensing Satellite (LCROSS) | USA | 2009 | Impact probe; polar volatile detection |
| Chang'e-2 | China | 2010 | Orbital remote sensing |
| Gravity Recovery and Interior Laboratory (GRAIL) | USA | 2011 | Orbital gravity mapping |
| Lunar Atmosphere and Dust Environment Explorer (LADEE) | USA | 2013 | Orbital exosphere and dust studies |
| Chang'e-3 | China | 2013 | Soft lander with rover: surface properties |

**Table 2.** Prioritized list of top-level lunar science concepts to be addressed by future lunar exploration [1]. Each of these top-level concepts may be further divided into individual scientific investigations (see table 3.1 of [1]), and appropriate mission architectures, instrumentation, and landing sites needed to address them can be identified [46-48].

| Science concept rank | Top-level description of lunar science concepts |
|---|---|
| 1 | The bombardment history of the inner solar system is uniquely revealed on the Moon |
| 2 | The structure and composition of the lunar interior provide fundamental information on the evolution of a differentiated planetary body |
| 3 | Key planetary processes are manifested in the diversity of lunar crustal rocks |
| 4 | The lunar poles are special environments that may bear witness to the volatile flux over the latter part of solar system history |
| 5 | Lunar volcanism provides a window into the thermal and compositional evolution of the Moon |
| 6 | The Moon is an accessible laboratory for studying the impact process on planetary scales |
| 7 | The Moon is a natural laboratory for regolith processes and weathering on anhydrous airless bodies |
| 8 | Processes involved with the atmosphere and dust environment of the Moon are accessible for scientific study while the environment remains in a pristine state |

# FIGURES

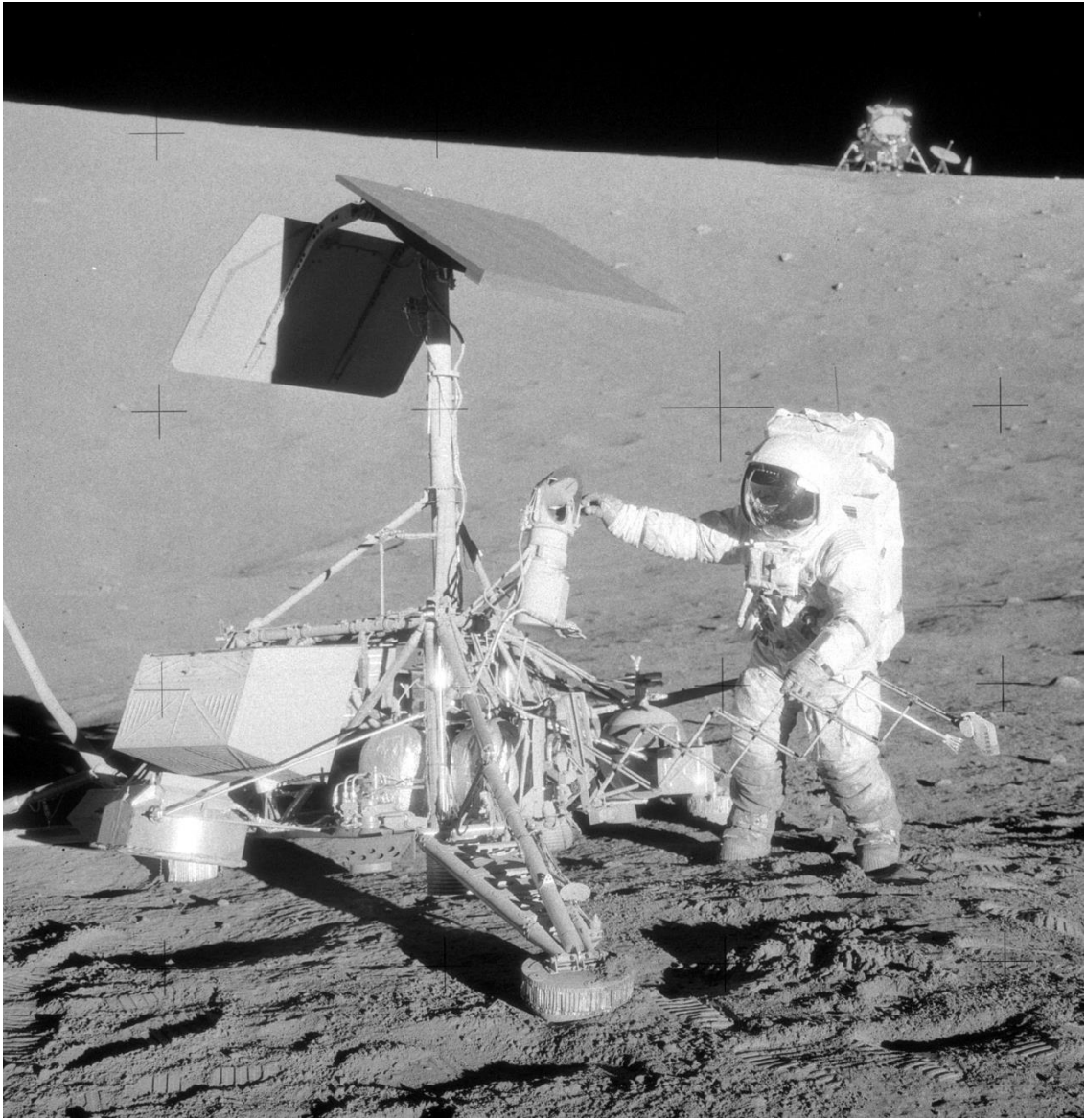

**Figure 1.** Charles 'Pete' Conrad, Commander of Apollo 12, stands next to Surveyor III in November 1969. Surveyor III had landed two and a half years earlier in April 1967. The Apollo 12 Lunar Module is in the background (NASA image AS12-48-7134).

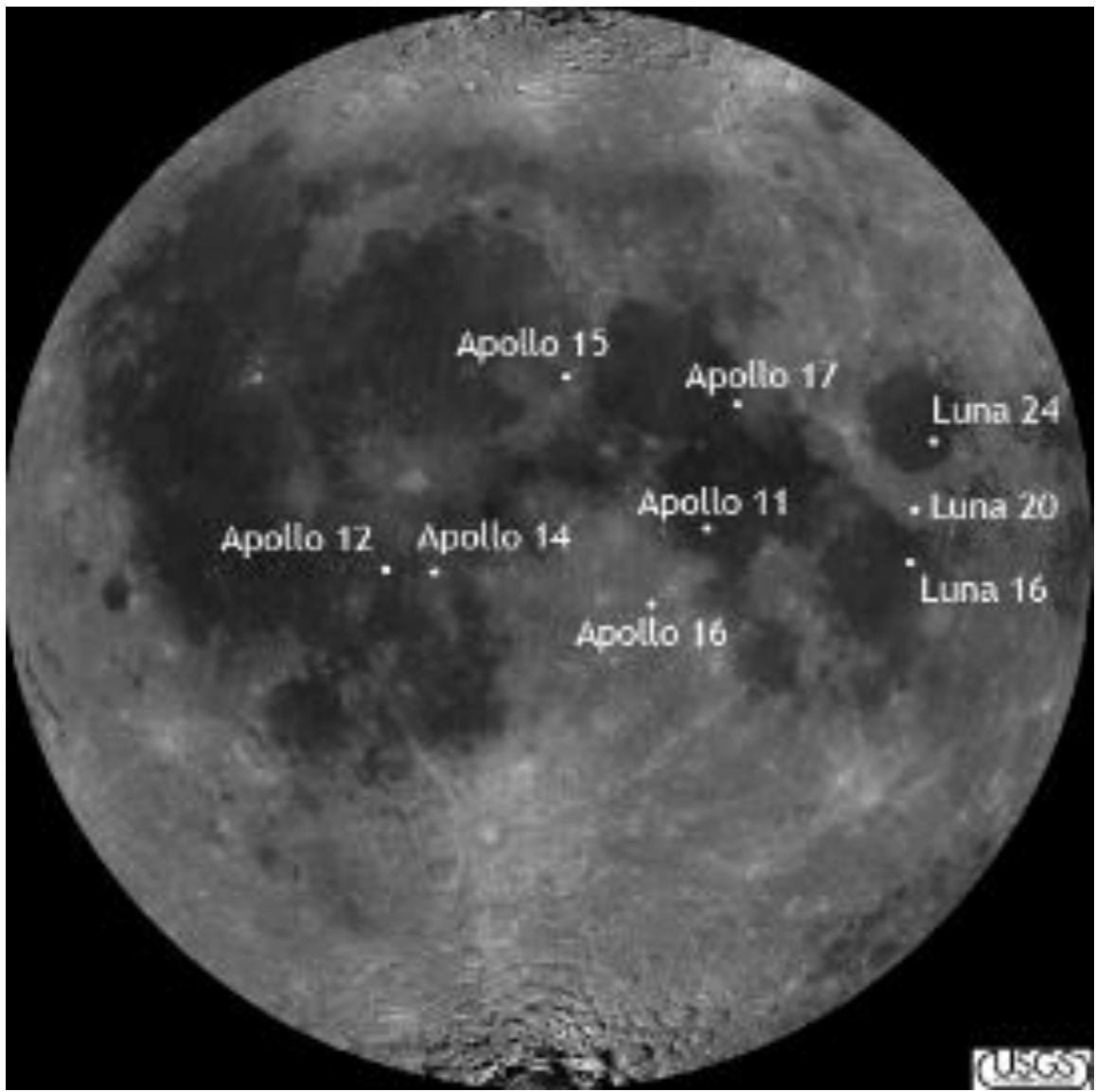

**Figure 2.** Nearside lunar mosaic constructed from Clementine 750 nm albedo data, with the landing sites of the six Apollo and three Luna sample return missions indicated. (USGS/K.H. Joy).

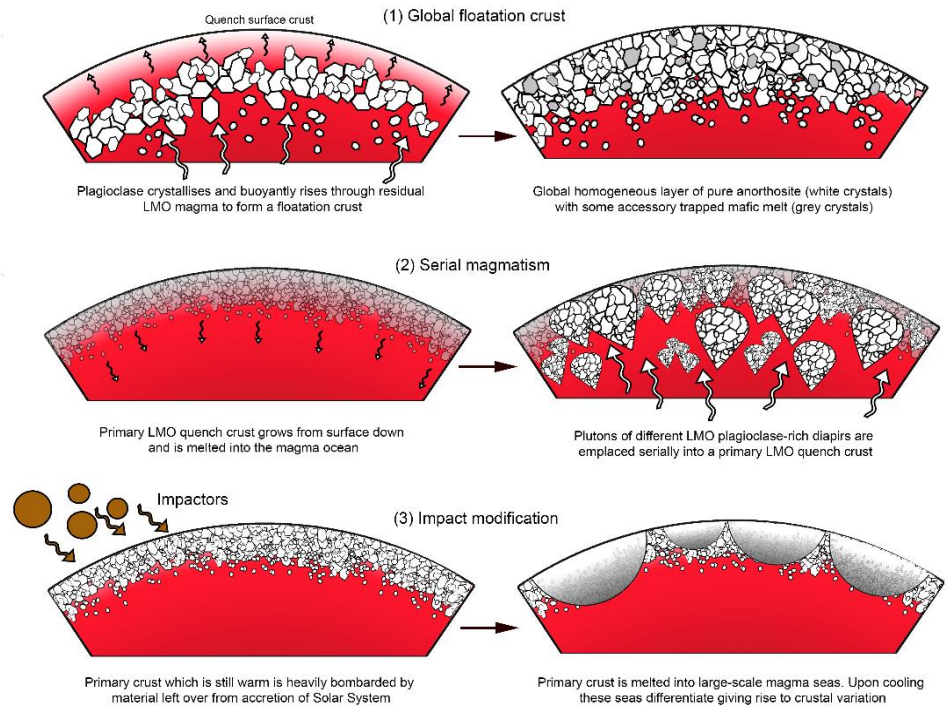

**Figure 3**. Schematic diagram of the differentiation of the Moon and the lunar magma ocean model. The top of the LMO, which was exposed to space, may have quenched and formed a thin surficial crust. After core separation the LMO crystallised mafic cumulates of olivine and pyroxene crystals. These cumulates sank into the interior to form the lunar mantle [25,49,67,70]. After ~80% of LMO crystallisation plagioclase became a liquidus phase and started to crystallise. Subsequent models of crust formation are debated [75,78]. Three possible models are illustrated here. (1) The traditionally accepted primary floatation crust model, whereby precipitated LMO plagioclase aggregated and floated on mass to the lunar surface forming a global anorthosite layer [70,82]. (2) Serial anorthositic magmatism, where crystallised LMO plagioclase rose as diapirs to form the crust. This type of model could either reflect primary crust formation or secondary crust formation if these diapirs were intruded into and replaced a pre-existing surficial quench crust [77,80]. (3) Modification of the lunar crust by widespread early bombardment, creating a series of magma seas that differentiate to form regionally variable crustal terranes [137]. The different models can be tested by (i) determining the ages of anorthosites [71,79] collected from different geological terrains (nearside, poles, farside) to understand if they all were formed at the same time (model 1) or at different times (models 2 or 3); (ii) determining and modelling the isotopic source regions and chemical variability of different anorthosites [71,77-79] to test if they have similar source compositions and melt evolutions (model 1 [70]) or variable source compositions and melt evolutions (models 2 or 3 [76-80]); (iii) determining global heat-flow and geophysical measurements to investigate is there a global KREEP (late-stage LMO melt) layer that would support the globally uniform LMO crystallisation and crust formation model (model 1); (iv) laboratory and modelling investigations of how lunar basin-forming impact melt sheets differentiate [139] to make predictions about crustal stratification, testable by remote sensing observations and direct sampling of lower crust material exposed in crater central peaks (diagram: K.H. Joy).

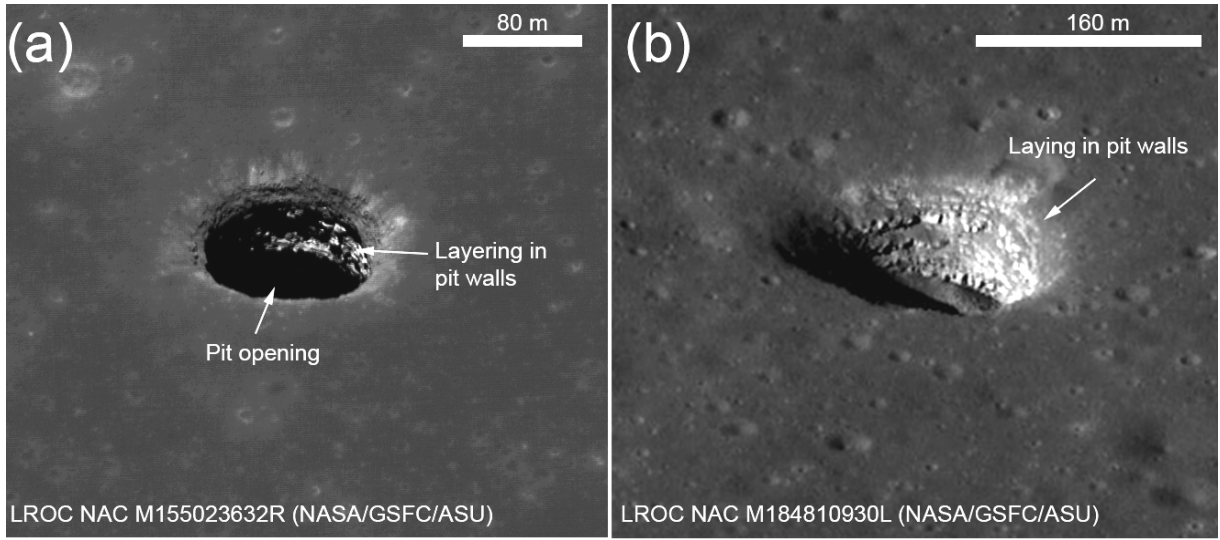

**Figure 4**. Oblique LROC Narrow Angle Camera (NAC) view of lunar pits with layered walls found in (a) Mare Tranquillitatis and (b) Mare Ingenii. The observed layering indicates that the maria are built up from multiple lava flows [140], and that sub-surface flows that have only been briefly exposed to the surface environment, and between which palaeoregoliths may be trapped, are likely to be common. (NASA/GSFC/Arizona State University).

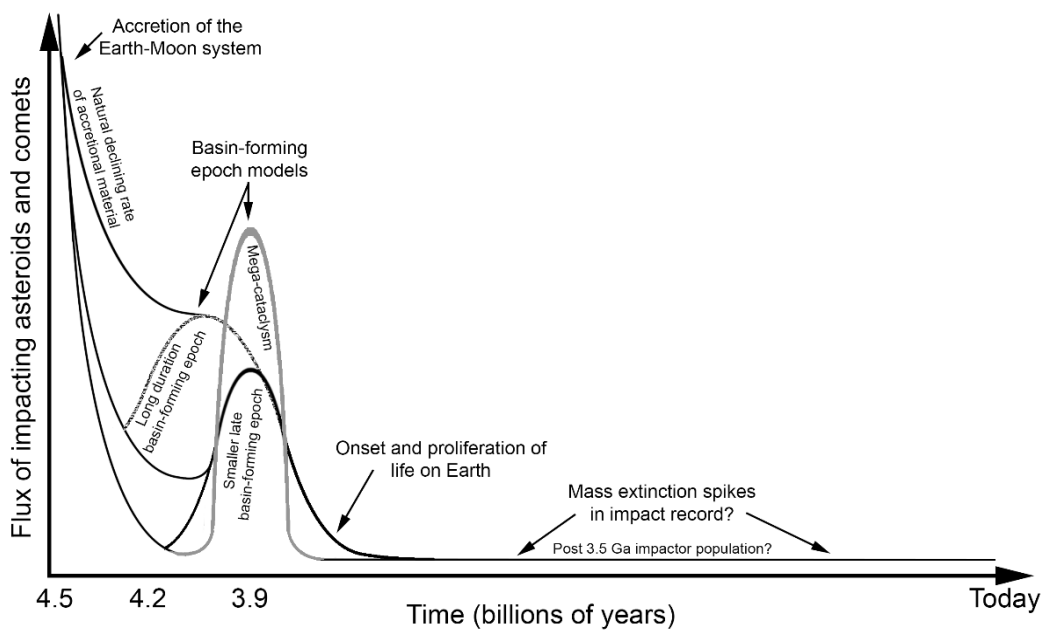

**Figure 5.** Schematic diagram illustrating various models of the impact cratering history of the Moon and the duration of the basin-forming epoch. Models range from no significant bombardment other than a declining primordial impact flux [87], a short cataclysmic spike in the impact record at ~3.9 Ga [87,100], to longer duration or saw-tooth model [96] of bombardment. Diagram modified from [87,94,95].

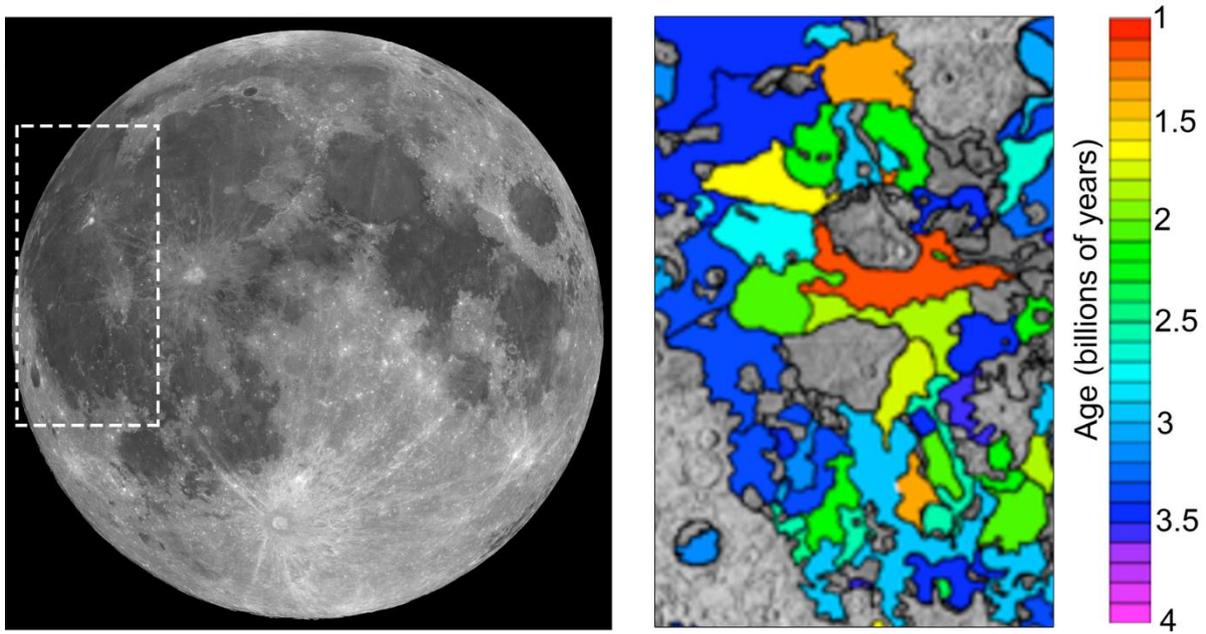

**Figure 6.** Left, albedo map of the near side of the Moon; dashed box represents region of Oceanus Procellarium mare basalts shown at right. Right, absolute model ages of lava flows in Oceanus Procellarum, as mapped by [106]. Sample return from one or more of these lava flows would verify these ages, with the benefits described in the text. (Image courtesy of Dr. H. Hiesinger; © AGU).

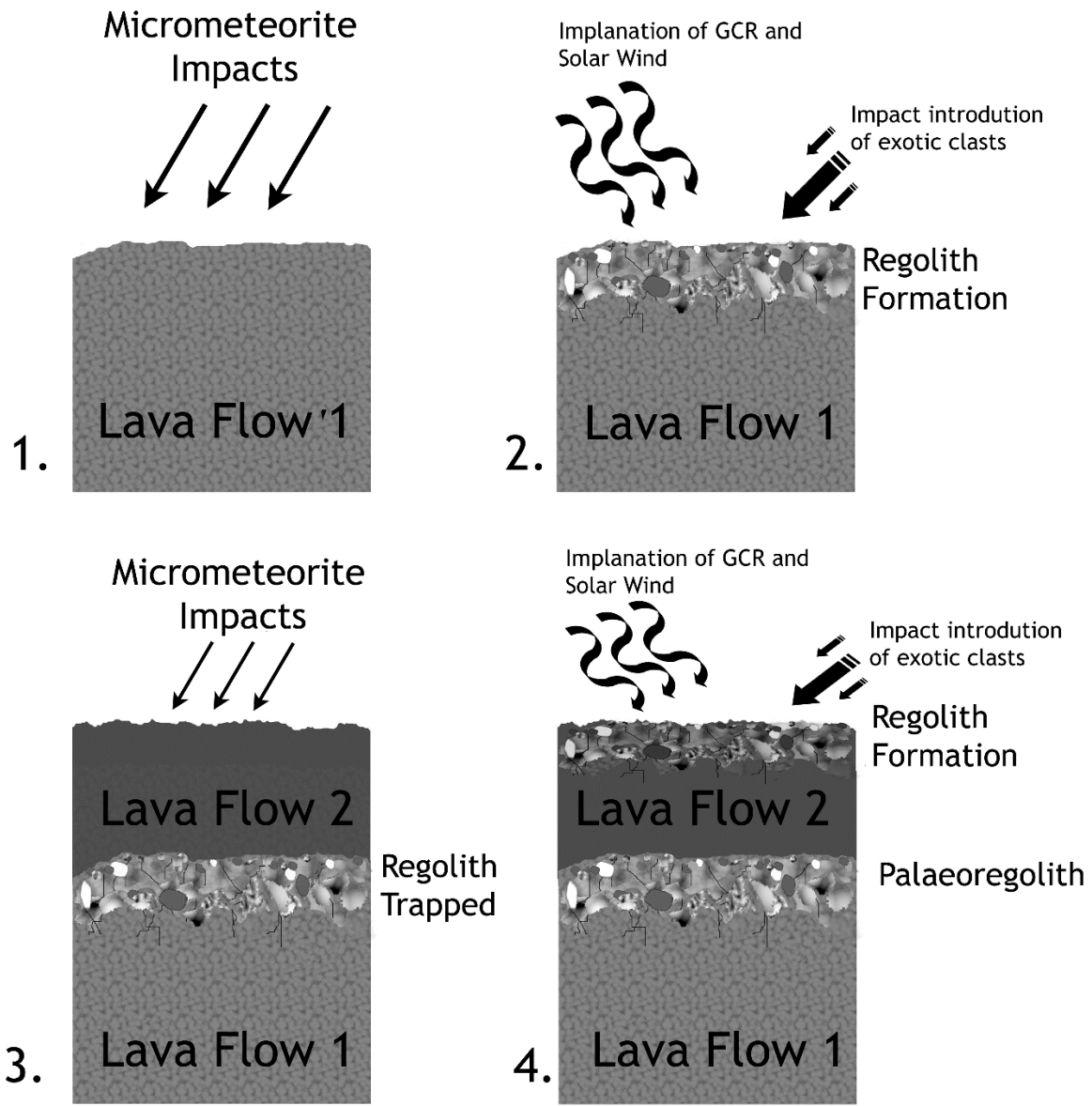

**Figure 7.** Schematic representation of the formation of a palaeoregolith layer [107]: (1) a new lava flow is emplaced, and meteorite impacts immediately begin to develop a surficial regolith; (2) solar wind particles, galactic cosmic ray particles and "exotic" material derived from elsewhere on the Moon (and perhaps elsewhere) are implanted; (3) the regolith layer, with its embedded historical record, is buried by a more recent lava flow, forming a palaeoregolith; (4) the process begins again on the upper surface. (© Royal Astronomical Society, reproduced with permission).